\documentclass{aa}
\usepackage{graphicx}
\usepackage{float}
\usepackage{multirow}
\usepackage{amsmath}
\usepackage{natbib}
\usepackage{enumitem}
\bibpunct{(}{)}{;}{a}{}{,} 

\newcommand{\bdel}{\mbox{\boldmath $\nabla$}}
\newcommand{\bdelp}{\mbox{\boldmath $\nabla_\perp$}}

\begin{document}
\title{Observations of apparent superslow wave propagation 
  \\ 
  in solar prominences} 


\author{J.O. Raes\inst{1}
  \and T. Van Doorsselaere\inst{1} 
     \and M. Baes\inst{2} \and A.N. Wright\inst{3}} 

\offprints{J.O. Raes, \email{jo.raes@kuleuven.be}}

\institute{Centre for mathematical Plasma Astrophysics, KU Leuven, Celestijnenlaan 200B, 3001 Leuven, Belgium \and Universiteit Gent, Krijgslaan 281-S9, Gent 9000, Belgium \and School of Mathematics and Statistics, University of St Andrews, St Andrews, KY16 9SS, UK}

\date{Date}

\abstract {Phase mixing of standing continuum Alfv\'en waves and/or continuum slow
waves in atmospheric magnetic structures such as coronal arcades can
create the apparent effect of a wave propagating across the magnetic
field.} {We observe a prominence with SDO/AIA on
2015 March 15 and find the presence of oscillatory motion. We aim to demonstrate that interpreting this motion as a magneto hydrodynamic (MHD) wave is faulty.  We also connect the
decrease of the apparent velocity over time with the phase mixing
process, which depends on the curvature of the magnetic field lines.} {By measuring the displacement of the prominence at different heights to calculate the
apparent velocity, we show that the propagation slows down over time, in accordance with the theoretical work of Kaneko et al. We also show that this propagation speed drops below what is to be expected for even slow MHD waves for those circumstances. We use a modified Kippenhahn-Schl\"uter prominence model to calculate the curvature of the magnetic field and fit our observations accordingly.} {Measuring three of the apparent waves, we get apparent velocities of 14, 8, and 4 km/s. Fitting a simple model for the magnetic field configuration, we obtain that the filament is located $103$ Mm below the magnetic centre. We also obtain that the scale of the magnetic field strength in the vertical direction plays no role in the concept of apparent superslow waves and that the moment of excitation of the waves happened roughly one oscillation period before the end of the eruption that excited the oscillation.} {Some of the observed phase velocities are lower than expected for slow modes for the circumstances, showing that they rather fit with the concept of apparent superslow propagation. A fit with our magnetic field model allows for inferring the magnetic geometry of the prominence.} 

\maketitle

\section{Introduction}
\begin{figure*}[t!]
\centering
  \begin{tabular}{@{}cccc@{}}
    \includegraphics[scale=0.5,trim={6cm 0 6cm 0}]{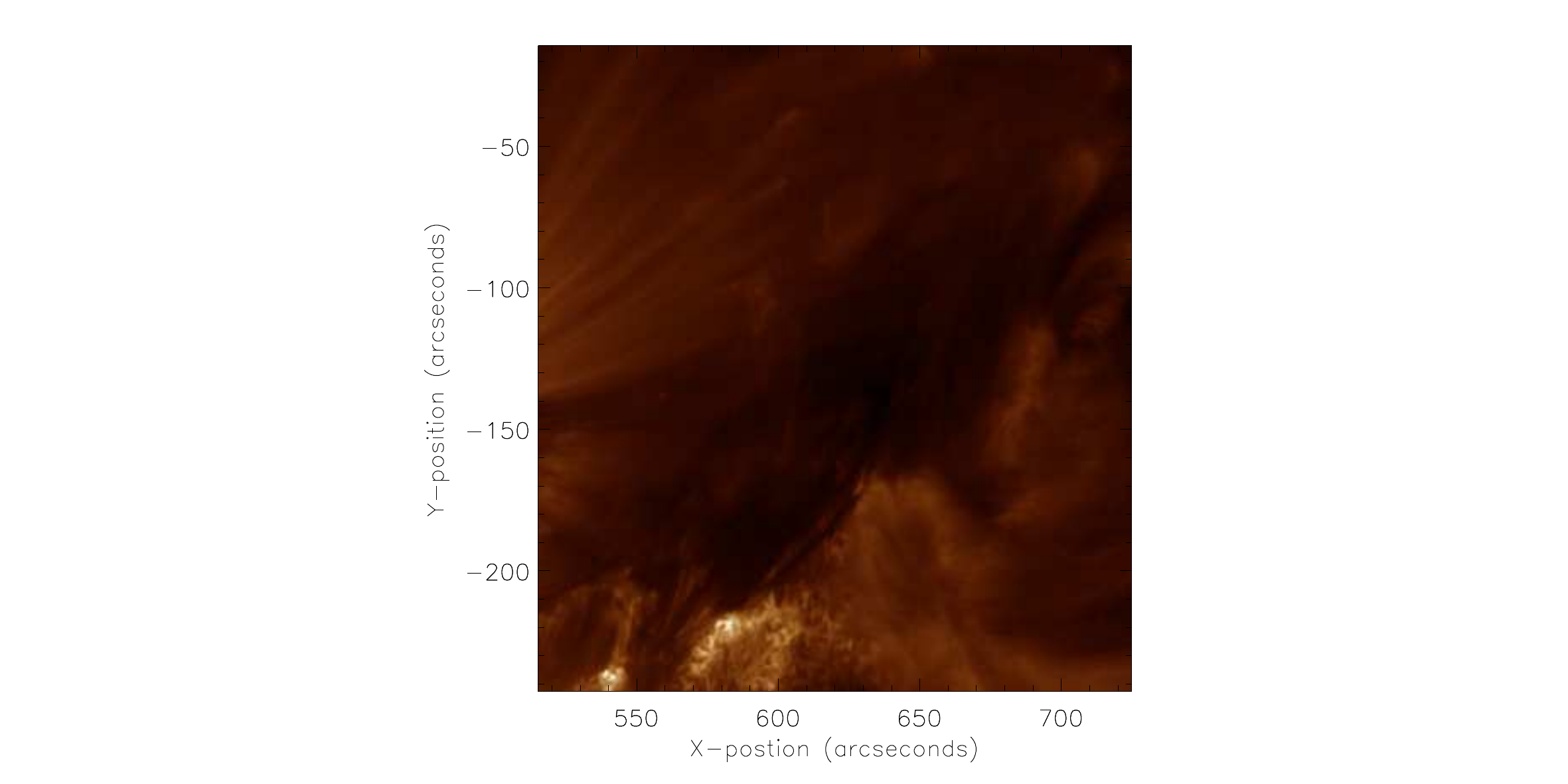} &
    \includegraphics[scale=0.5,trim={6cm 0 6cm 0}]{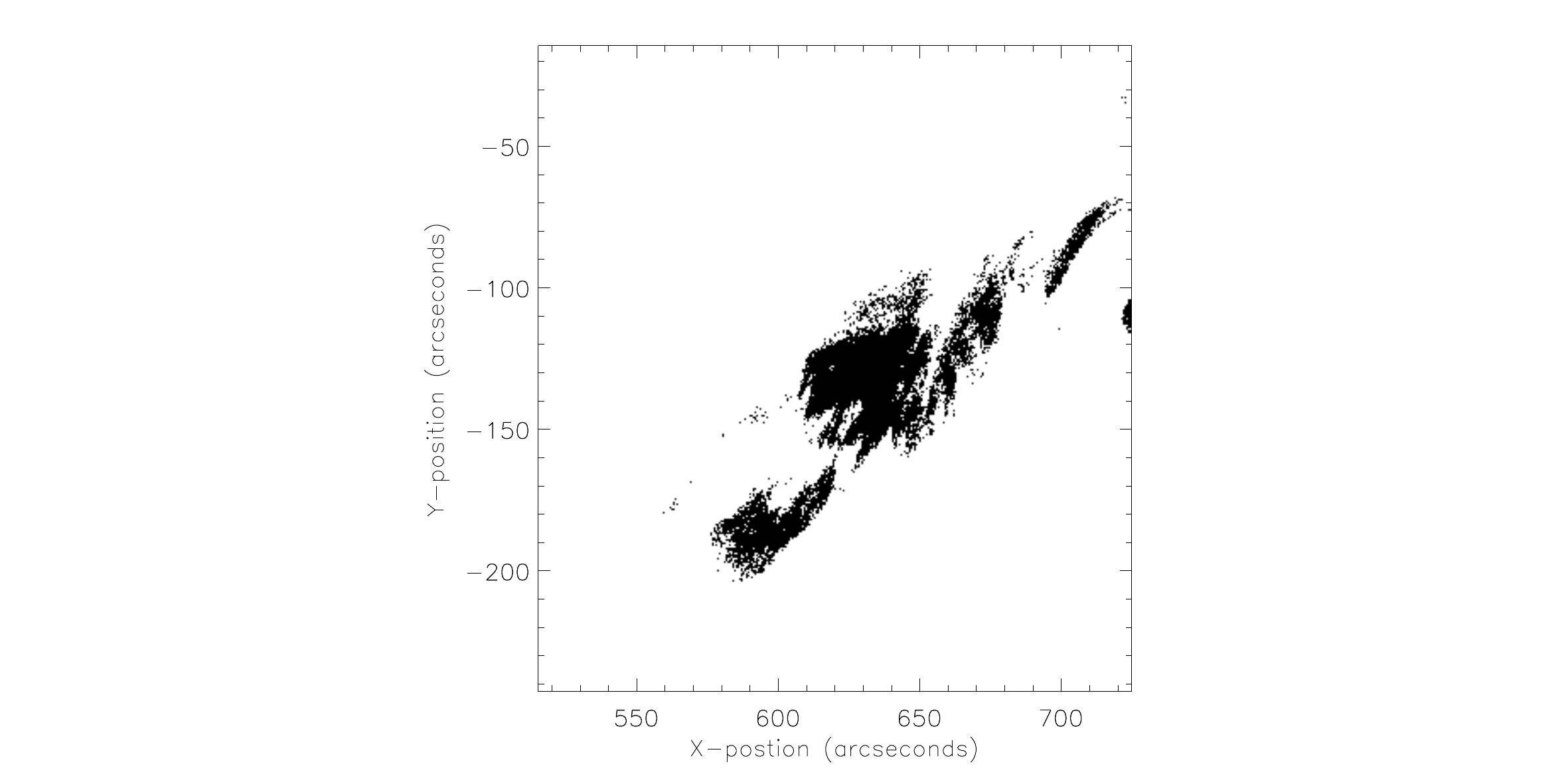} &
    \includegraphics[scale=0.52,trim={0 0 0 0}]{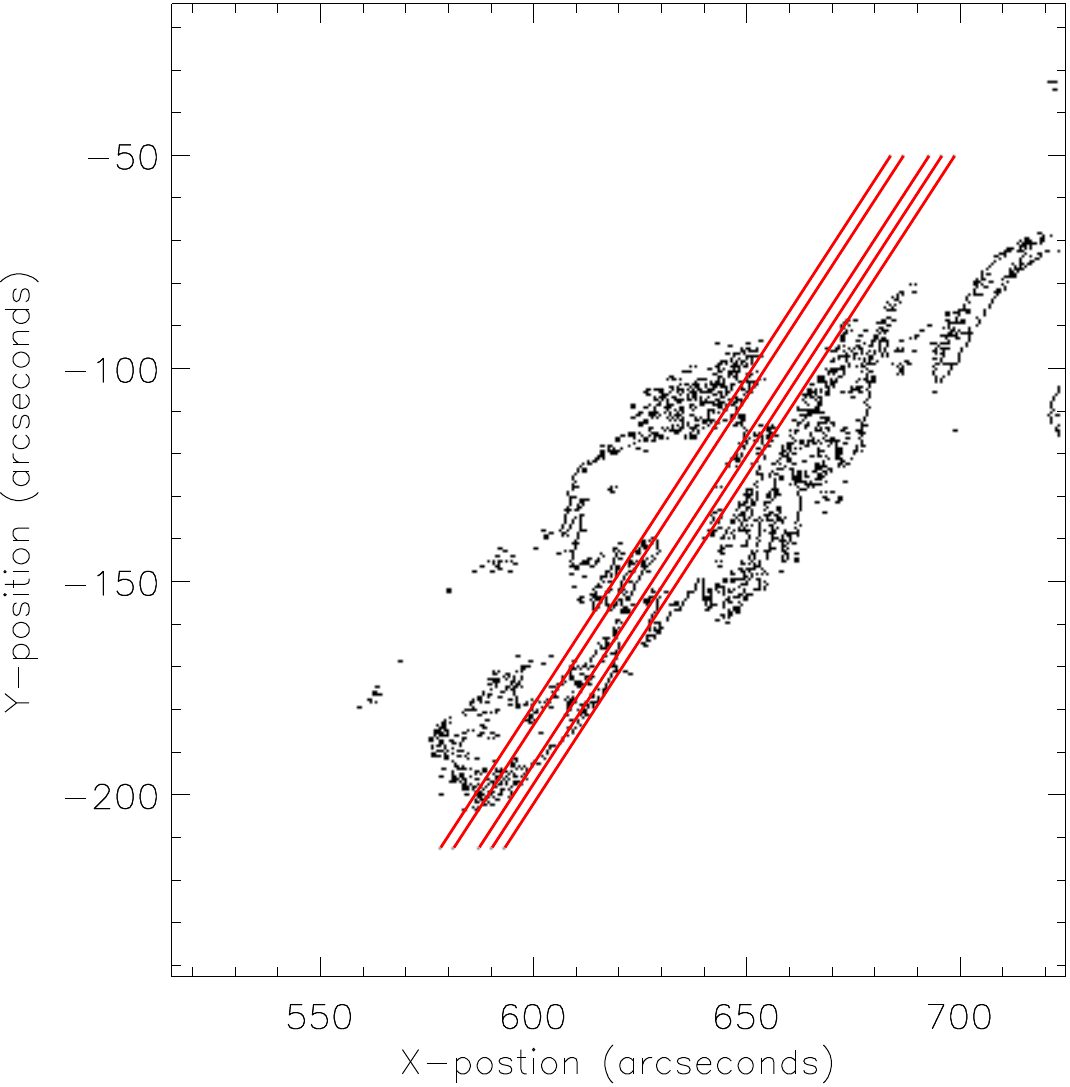}   \\
  \end{tabular}
  \caption{Prominence observed by AIA on 2015 March 15 at 03:10 in the 193 \r{A} channel. The left panel shows the data in a non-manipulated format. The darker area denotes the filament and the brighter area is the solar background. The temporal evolution as seen in AIA 193 channel is available online. The middle panel shows the data set after a first manipulation by using a cut-off value. In this format, the prominence can be seen as a black shape against a white background. Compared to the non-manipulated data, the prominence can be seen much more clearly. The right panel shows the data after a final manipulation: in this format only the prominence edges are displayed. The red lines over the prominence are the different slices at which the displacement measurements are taken.}
  \label{prominencefigures}
\end{figure*}
Solar prominences are huge magnetic structures consisting of large amounts of solar plasma suspended in the solar corona. Compared to their coronal surroundings typically they are roughly 100 times cooler and denser, with temperatures up to $10^4$ K and electron densities of $10^9$ to $10^{11}$ cm$^{-3}$ \citep[for a review, see][]{ 2010SSRv..151..243L, 2010SSRv..151..333M}. 
\\
Oscillatory motion in prominences and other coronal structures, such as loops and plumes, have been of scientific interest for a while now. While they have been observed by H$\alpha$ spectrograms as early as the 1930s \citep{test}, theoretical studies on the subject long predate observational evidence owing to a potential link with the coronal heating problem \citep{1992A&A...256..264J, 2010SSRv..151..333M, 2003A&A...402..781D}. The ability to study oscillations has advanced drastically over the years thanks to improved observational methods, such as two-dimensional spectrographs and image stabilisers, and analysis tools, such as wavelet transforms \citep{2002SoPh..206...45O}. At the moment the main method to detect these motions is through the periodic Doppler shifts of spectral lines or observed displacements. For prominences these observations have shown that the oscillations are mostly localised and undergo strong damping over time \citep{2010SSRv..151..333M}. Depending on the amplitude of the oscillations they can be divided into two groups: small-amplitude oscillations and large-amplitude oscillations \citep{2012LRSP....9....2A}.  \\
Small-amplitude oscillations can be distinguished due to containing one of three aspects: 
\begin{enumerate}[label=(\roman*)]
    \item Only a restricted part of the prominence is subjected to the oscillation
    \item The amplitude of the oscillation is rather small
    \item The relation to flare activity is usually non-existent.
\end{enumerate}
Aside from the size of the amplitude, the most important characteristic of large-amplitude osculations is the fact that the entirety of the prominence undergoes the movement, in which displacements from the equilibrium position ranging from a few thousand to a few ten thousand km. For a long time it was believed that large-amplitude oscillations were only caused by the collision of the filament with a Moreton wave (a flare-associated wave that propagates in the chromosphere) \citep{1960AJ.....65U.494M, 2004ApJ...608.1124O}. More recent observations however exhibit the presence of large-amplitude oscillations without the presence of a remote flare and thus without the accompanying Moreton wave; other triggering events could be magnetic reconnection between a filament barb and a nearby emerging flux \citep{2007SoPh..246...89I} or a subflare \citep{2003ApJ...584L.103J}.
A handful of models have been introduced to explain large-amplitude oscillations. One of the earliest is the Kleczek \& Kuperus model \citep{1969SoPh....6...72K}. In this model, the filament is represented as a slab with the magnetic field running along it. Oscillatory motion is perpendicular to the main axis of the slab and magnetic tension plays the role of the restoring force. \citet{2003ApJ...584L.103J} used this model as a basis for the interpretation of three filament observations on 2001 October 24, 2002 March 20, and 2002 March 22. One of the main conclusions in this work is that the direction of the observed oscillations conflicts with that of the Kleczek \& Kuperus model. The observations seem to show that the displacements are mostly oriented along the filament axis, but the magnetic tension drives transverse motions, perpendicular to the actual oscillation \citep{2006SoPh..236...97J}. \citet{2007A&A...471..295V} proposed a model of a flux rope geometry with oscillations analogous to a longitudinal-mode standing wave on a spring fixed at both ends. This model however also predicts motions that are oriented perpendicular to the local magnetic field, which is a phenomenon not found in observations. One of the more recent works on the topic is by \citet{2016arXiv160503376R}. In their model, the oscillation consists of the unified motion of multiple cool, dense threads along the magnetic field. A nearby energetic event, such as a flare, subflare, or microflare, is taken as the triggering event. The restoring force is the projected gravity in the flux tube dips and the oscillation is damped  by mass accretion of the threads \citep{2012ApJ...750L...1L, 2016arXiv160503376R}.\\
In recent years, there have been a number of numerical simulations regarding prominence oscillations. \citet{2012A&A...542A..52Z} used Hinode high resolution observations and attempted to reproduce the observed damped oscillations by performing a one-dimensional hydrodynamical numerical simulation. In their results they show that the oscillation period derived from the simulation closely matches the observed one and their findings seem to support that the projected gravity is the restoring force, as mentioned by \citet{2012ApJ...750L...1L}. \citet{2013ApJ...778...49T}  calculated two-dimensional numerical models that connect the magnetic field to the photosphere and include an overlying arcade. Oscillatory motion is simulated by injecting mass into the equilibrium state of the system. These authors found that vertical oscillations are always stable for their equilibrium parameters when there is no perpendicular propagation. On the other hand, Longitudinal oscillations, which are mainly related to slow magnetoacoustic-gravity waves, can become unstable because that they are more strongly affected by gravity. This two-dimensional model was later expanded to a three-dimensional model \citep{2015ApJ...799...94T, 2016ApJ...820..125T} while using the same concepts as in the two-dimensional model. In these simulations the main objective was to tie the time evolution of the prominence to the different parameters of the configuration, where plasma $\beta$ is one of the more critical parameters.
\citet{2016A&A...590A.120K} developed an analytical model for transverse oscillations. In this model, they account for both the magnetic dip and mirror current, which is a current located below the prominence that is generated by the conductive properties of the photosphere. In their results, they find the properties of vertical and horizontal oscillations and show that the system is in fact stable when the force of the mirror current is accounted for.\\
When cross-field propagation is observed in solar filaments, it is usually attributed to magnetosonic MHD waves. \citet{2015ApJ...812..121K} and \citet{2013ApJ...777..108S}, however, have shown that this may in fact be faulty. Magnetic surfaces in the prominence can contain trapped continuum Alfv\'en waves. In ideal MHD, a single flux surface oscillates at its own frequency without any influence on or from neighbouring flux surfaces (negligible effect in non-ideal MHD). Depending on the variation of the frequency through the filament, an illusionary effect of a propagating wave can be created that can be confused with an MHD wave. The simulations of \citet{2015ApJ...806..115K} have shown that these apparent waves slow down over time with propagation velocities that are lower than fast and even slow modes. \\
While these observational and theoretical works have been around for some time now, applying seismology to solar oscillations is a more recent application \citep{2014IAUS..300...30B}. Seismology entails the analysis of oscillation or wave properties to study the conditions of the medium through which they travel, which can be applied to a number of different fields. 
Solar atmospheric seismology, while introduced as early as 1970 \citep{1970A&A.....9..159R, 1970PASJ...22..341U, 1984ApJ...279..857R} was only fully realised since the late 1990s \citep{1999Sci...285..862N, 2001A&A...372L..53N, 2002A&A...394L..39G, 2007A&A...463..333A, 2005ApJ...624L..57A, 2011ApJ...727L..32V, 2016IAUS..320..202W}. \\
One can use MHD seismology to determine physical parameters of plasma structures, such as coronal magnetic field, transport coefficients and heating function. When considering prominence seismology, both large-amplitude oscillations and small-amplitude oscillations can be used as an observational tool \citep{2014IAUS..300...30B}. Related observations have been reported in a magnetospheric context, where phase motion has been observed at the ionospheric footpoints of field lines supporting Alfv\'en waves \citep[see][for a review of the observations and theory used to interpret them]{2006GMS...169...51W}. \\
The aim of this paper is to add more observational evidence for apparent superslow wave propagation in prominences, by showing that the oscillatory motion observed in a filament on 2015 March 15 can be attributed to this concept. This will be carried out using observations from the Atmospheric Imaging Assembly (AIA) aboard the Solar Dynamics Observatory (SDO) to determine the apparent phase velocity of the observed movement. Section 2 explains the data reduction in more details. The calculation of the apparent phase velocity and subsequent results can be found in section 3. In section 4 we connect the phase mixing process with the decrease of apparent velocity over time using a model for the prominence magnetic field with the aim of using the superslow waves for seismology. We fit the model to the data to infer the magnetic configuration. Our conclusions are formulated in section 5. 

\section{Observations}
The SDO/AIA observations of the prominence were carried out from 00:00 UT - 10:00 UT on 2015 March 15. This case was chosen by visual inspection. It is observed at 600 arcsec solar west, 130 arcsec solar south, which is approximately half a solar radius from the solar centre. The location is near active region 12297. The prominence is clearly visible in SDO/AIA filters, mainly in wavelengths of 193 and 304 \r{A} . The central panel of Figure \ref{prominencefigures} shows the prominence in AIA 193 \r{A}, where it can be seen as a dark region against the brighter solar disk.
\subsection{Observed oscillatory motion}
The AIA 193 \r{A} observations clearly show an eruptive event near active region 12297 from 00:00 UT - 03:00 UT. Consequently, oscillatory motion can be seen until approximately 10:00 UT, but it is most outspoken from 03:10 UT - 06:20 UT. The exact moment when the eruption excites the oscillatory motion cannot be found in the observations. In section \ref{fitobserv} we show that the oscillatory motion starts around 02:15 UT, but the eruption itself visually blocks any sign of this. In order to verify whether these motions are MHD waves or not, we determine the (apparent) phase velocity of the wave. This can be achieved with accurate measurements of the wave amplitude over time at different heights of the prominence. For this we need to determine the location of the prominence edges. 

\subsection{Data reduction}
The data from SDO AIA has a cadence of 12 s. Each frame consists out of 4096 by 4096 data points. We used a total of 2990 frames, covering a time span from 00:00 UT to 10:00 UT on 2015 March 15, from the 193 \r{A} channel. Most of these data are obsolete however, as only a small part of the solar surface contains the prominence and the oscillatory motion is only observed in a smaller time interval. Cutting the unnecessary parts we get 980 frames consisting of 349 by 380 data points. A first step to locating the edges of the prominence is to introduce an intensity cut-off in these data. By changing all data values below this cut-off value to zero and all data values above or equal to this cut-off to one, we can find a clearer picture for which part of the image is the prominence. Through a process of trial and error we find that a cut-off value of $30200$ DN yields the best result. We now have a data set where a data point with a value of 0 belongs to the prominence and a data point with a value of 1 denotes the solar background. The left panel of Figure \ref{prominencefigures} shows an image of these data format: a black blob against a white background. As can be seen, these data show the prominence edges very clearly. \\
The location of the edges can now be found by comparing neighbouring data points in the vertical direction. When two neighbouring data points have the same value (either 0 or 1), we regard them as belonging to same medium (either prominence of solar disk). When two neighbouring data points have different values, these two points form a transition from solar disk to prominence or vice versa. In other words, where the difference between two sequential data points is non-zero, we are looking at the prominence edge. This way, we create a new dataset in which a value of 1 denotes the location of an edge and a value of 0 denotes the rest. An image of this dataset can be seen in the right panel of Figure \ref{prominencefigures}.\\
We take a set of slices across the filament to obtain the displacement of the oscillatory motion at different heights. For each slice we determine the intersection of the slice with the northern edge of the filament, for each frame. Subsequently we calculate the distance between this intersection and a fixed reference point on the slice. The exact location of this reference point is not of great importance: we want to know when the displacement is maximal, not its exact value. Measuring the projected distance between the slices and setting the lowest slice at height 0, Figure \ref{height0} shows the displacement over time at projected height 1530 km. The effect of projection due to the angle of the observation is not taken into account, such that all speeds and distances measured in this article are projected distances.
\section{Results}
\begin{figure}[h]
\begin{center}
\includegraphics[scale=0.45]{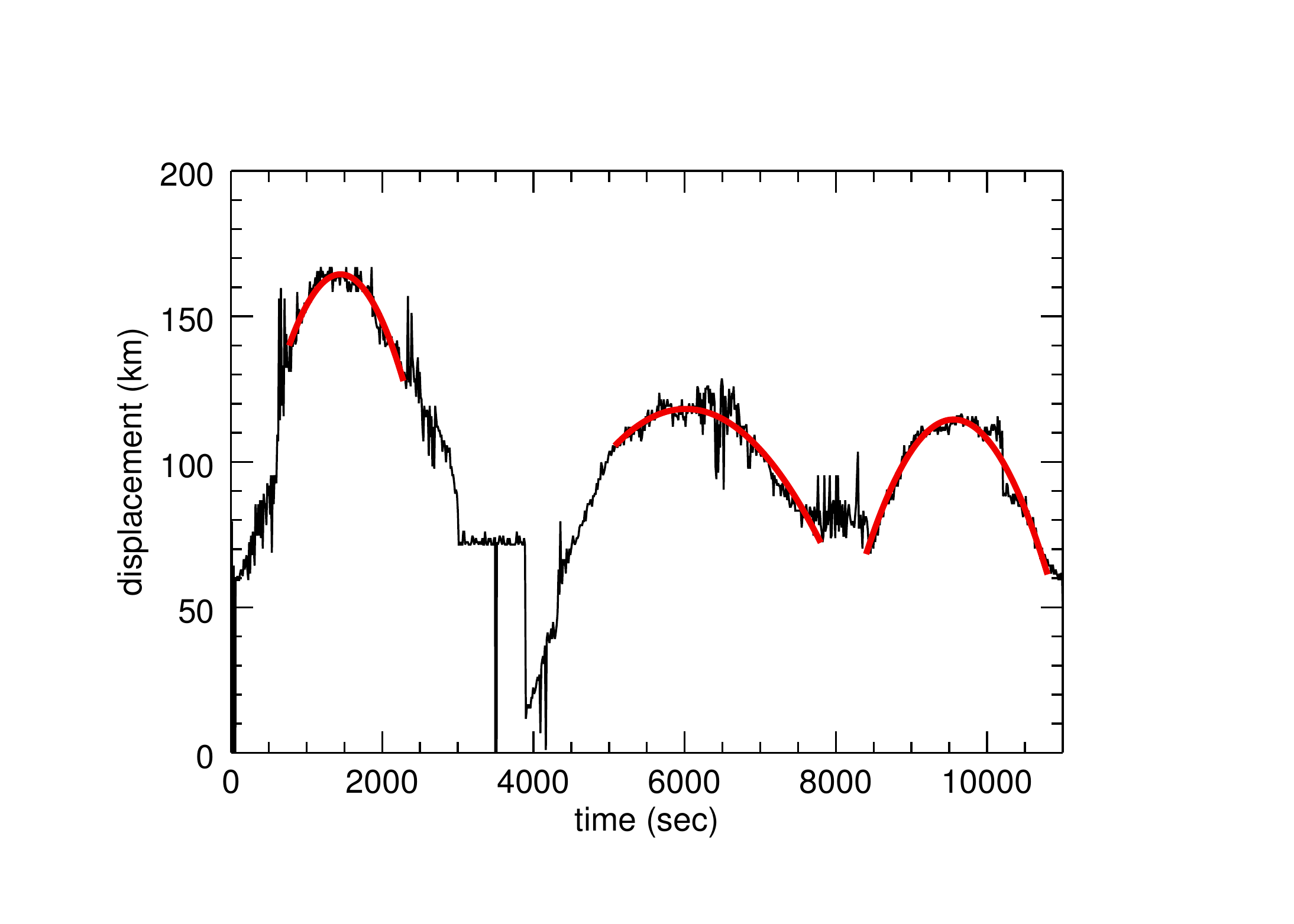}
\caption{Displacement measurements of the filament at height 0. The red lines at each peak are fitted parabola to find the time of the maxima.}
\label{height0}
\end{center}
\end{figure}

\begin{figure}[h]
\begin{center}
\includegraphics[scale=0.5, trim={0.9cm 0 0 0}]{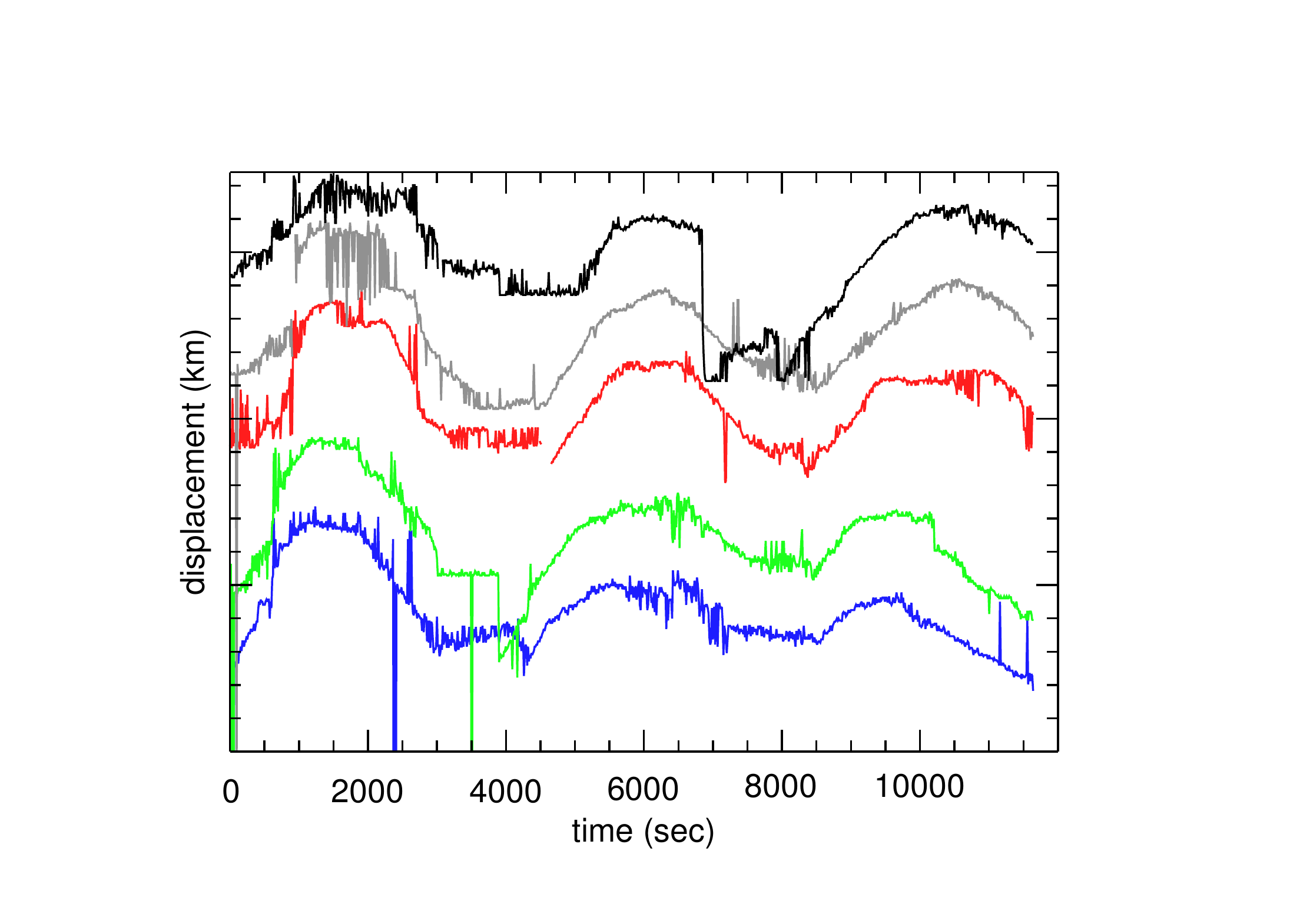} 
\caption{Displacement measurements of the filament at all heights. The blue curve is at a height of 0 km, the green curve at height 2175 km, the red curve at height 6527 km, the grey curve at height 8703 km, and the black curve at height 10879 km; these are projected heights. The time difference between maxima for subsequent slices is smallest during the first maximum, larger for the second maximum, and largest for the third maximum.}
\label{allplots}
\end{center}
\end{figure}

\begin{figure}[h]
\begin{center}
\includegraphics[scale=0.45, trim={0 1.5cm 0 0}]{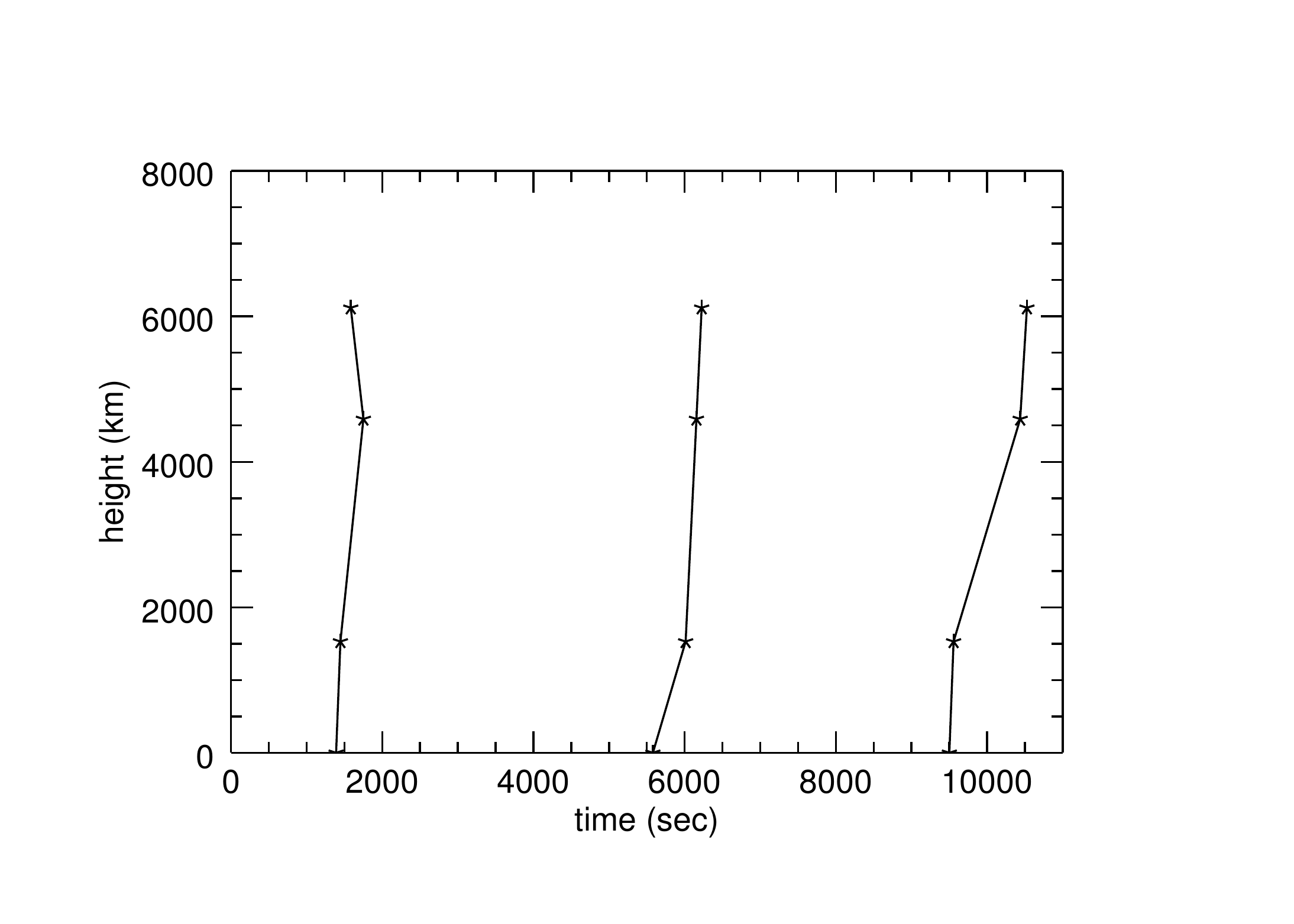} 
\caption{Plot of the times of maximal displacement for the different projected heights. The slopes indicate that the velocity decreases over time for each peak.}
\label{peaktimesplot}
\end{center}
\end{figure}

To get an initial of the apparent phase velocity, we gather the amplitudes of the slices at each height into a single plot, as shown in Figure \ref{allplots}. In this figure a constant offset proportional to the height is added to the displacement of each subsequent slice. For this reason, no numerical values are given on the vertical axis, as otherwise the highest slice would seem to have a much higher amplitude than the lowest. We note that there are instances with a lot of noise in the data. This is because the intensity profile does not always follow a smooth shape, making automated edge detection noisy. Going over each data frame separately and manually selecting the edge would have resulted in much less noise. However, we used a systematic approach to avoid any bias. Luckily, the noise is minimal around the peaks of each displacement graph, so we can still determine the moment of maximum displacement accurate enough. This is also the reason why we use the northern edge of the filament, where there is less noise.
Figure \ref{allplots} gives an interpretive grasp on the apparent velocity; as the apparent wave moves through the filament, the peak time occurs later for each subsequent slice \citep{2013ApJ...777..108S}. We thus need the times of each local maximum of the amplitudes for each projected height to find the apparent phase velocity. As the highest slice is too close to the boundary of the prominence, its results are unreliable and are not used for further analysis. To find the maxima of the other slices, we fit a parabola to each peak of every slice as can be seen by the red lines for the slice of height 1530 km in Figure \ref{height0}. For the fits we took an interval of about 1200 s for the first peak and 2400 s for the second and third peaks, all centred on a rough estimate of the maximum. The times can be found in table \ref{peaktimes} and are plotted in Figure \ref{peaktimesplot}. The time of 0 s corresponds with the end of the eruption, which is 2015 March 03:10 UT. 
\begin {table}[h]
\begin{center}
\begin{tabular}{ c  c c c }
  \hline   \hline                     
  Height (km) & Peak 1 (sec) & Peak 2 (sec) & Peak 3 (sec)\\
  \hline
  0 & 1390& 5580 & 9501\\
  1530 & 1447 & 6013 & 9557\\
  4589 & 1751 & 6156 & 10438\\
  6118 & 1584 & 6225 & 10527\\
  \hline
\end{tabular}
\caption {Overview of the times of maximal amplitude for each slice.}
\label{peaktimes} 
\end{center}
\end {table}

\begin{figure}[h]
\begin{center}
\includegraphics[scale=0.45, trim={0 0 0 0}]{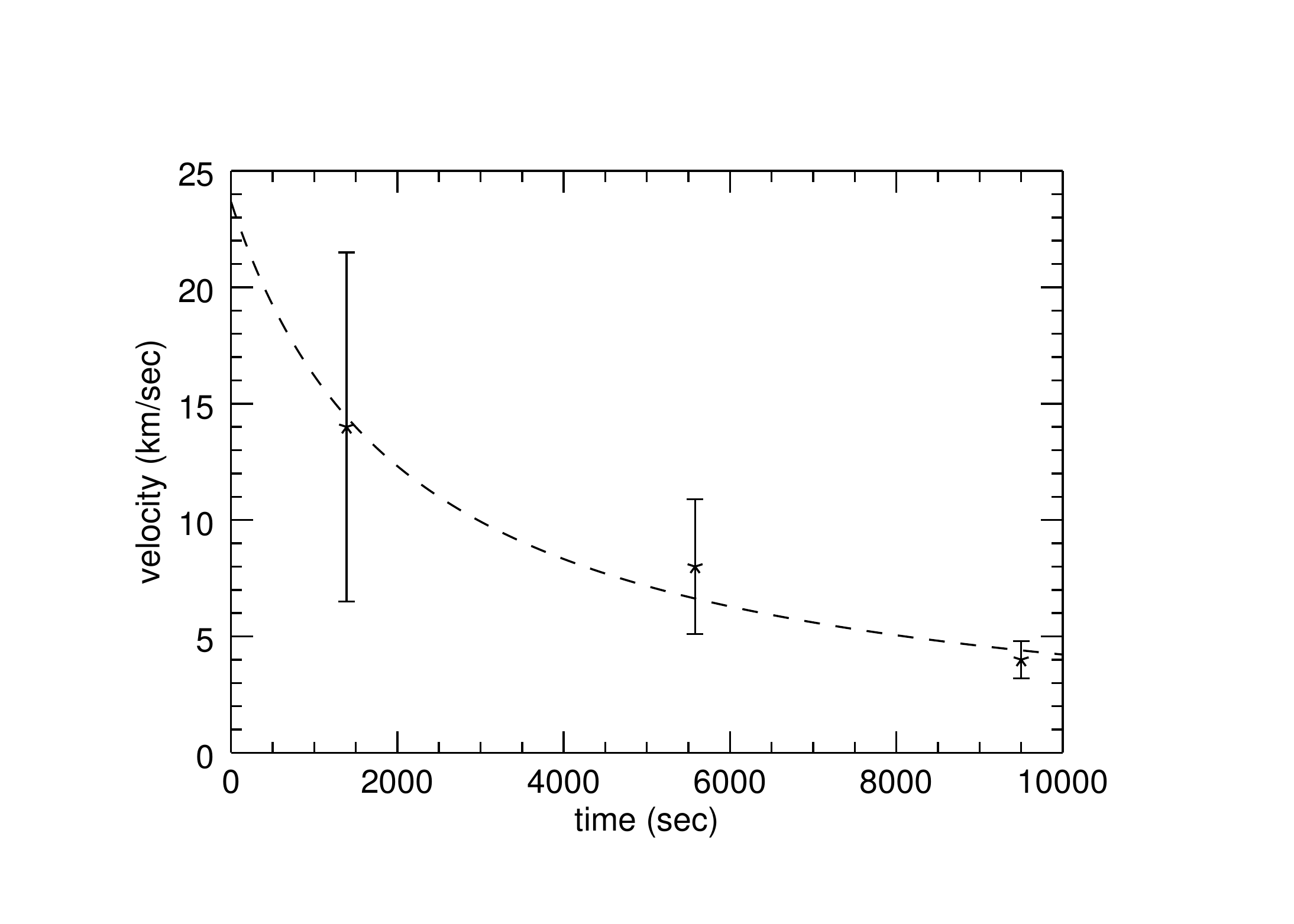} 
\caption{Overview of the apparent phase velocities over time. The time for each velocity corresponds with the time when the first maximal amplitude was reached for that wave at height 0. The dotted line is the fitted phase velocity using eq. \ref{finalvelocity}.}
\label{velocitieshigh}
\end{center}
\end{figure}

We can easily get the velocities from each peak by interpolating these points. This gives us 14, 8, and 4 km/s for the three peaks, respectively. Using the same moment in time for $t=0$ as defined above, gives an evolution of phase velocity over time as shown in Figure \ref{velocitieshigh}.
We first notice the decline of the apparent phase velocity over time, as expected from the theoretical work by \citet{2015ApJ...812..121K} and the numerical simulations by \citet{2015ApJ...806..115K}. This indicates that we are dealing with apparent superslow waves instead of MHD waves. When looking at the expected values of phase velocity for MHD waves, one would presume approximately 20 km/s or higher for solar prominences \citep{2010SSRv..151..333M}. While the first observed apparent phase velocity of 14 km/s could still be considered ambiguous, the values of 8 and 4 km/s are vastly lower than expected speed for even the slow mode. This suggests that interpreting the observed oscillatory motion of the prominence as MHD waves is not correct.
\section{Seismology}
In this section we attempt to combine the works of \citet{2015ApJ...812..121K} and \citet{2012ApJ...746...30L} to fit our observations of the superslow propagation to the change of frequency using a magnetic field model.

\subsection{Phase mixing and phase motion}

There are many instances in MHD where individual field lines exhibit natural oscillations along their length that are essentially decoupled from neighbouring field lines. Examples include Alfv\'en waves, slow modes, and the gravity-driven sloshing modes considered by \citet{2012ApJ...750L...1L}. Since the frequency of the oscillation varies from one field line to another, considering a set of field lines in a smoothly varying medium leads to a continuum of permitted natural frequencies. 

This has been studied previously in two-dimensional systems where the flux function is a natural coordinate. For example, \citet{1999JGR...10410159W} show how the scales and motion phase structures in standing Alfv\'en waves may be predicted. \citet{2015ApJ...806..115K} provide a similar analysis for interpreting coronal Alfv\'en waves in a simulation. In this subsection we indicate how the ideas of phase mixing and phase motion may be generalised to a three-dimensional system.~\vspace{0.5cm}


To facilitate analysis it is natural to introduce a field-aligned coordinate system in which the two perpendicular directions are identified with coordinates $\alpha$ and $\beta$, for example Euler potentials. The following analysis applies when the continuum frequency may be denoted by $\omega_c (\alpha ,\beta )$. Since these coordinates are constant on a field line it also guarantees that the frequency is the same everywhere along a particular field line. The coordinates are completed by a field aligned coordinate ($\gamma$). Whilst it may be difficult to define ($\alpha ,\beta ,\gamma $) as orthogonal coordinates globally in certain cases, such as when there is a field aligned equilibrium current, there is no problem if we are considering a smaller subdomain in the vicinity of a chosen field line as we do here.

We begin with considering the natural undamped continuum oscillations. Assuming the system to have been excited at $t=0$, the subsequent state (for $t> 0$) of the perturbation quantity $\xi$ may be represented by
\begin{eqnarray}
\xi (\alpha ,\beta ,\gamma, t) = a(\alpha ,\beta ,\gamma )\exp \left[-i\omega_c(\alpha,\beta)t\right], \label{perturbationquanity}
\end{eqnarray}
where the complex coefficient $a(\alpha ,\beta ,\gamma )$ is determined from initial conditions. The quantity $\xi$ could represent any leading order continuum field, such as a component of velocity, magnetic field, displacement, etc., associated with the natural oscillation. For some systems there may be several harmonics present, in which case these should be summed over. For simplicity we assume that there is only a single mode that dominates the behaviour. Depending upon the system considered, the above expression could be an exact representation or one that is asymptotically valid. 

The field aligned eigenmode structure is contained in the coefficient $a$, as is the initial cross-field variation of $\xi$. In a one-dimensional system,  \citet{1995JGR...10019441M} showed how the solution develops increasingly small scales ($\propto 1/t$) in the perpendicular direction owing to phase mixing, which is a property of a time-dependent evolution. Here we generalise their results \citep[and those of][]{1999JGR...10410159W} to three dimensions. Taking $\mbox{\boldmath $\nabla$}_\perp$ of Eq. \ref{perturbationquanity} gives
\begin{eqnarray}
\bdelp\xi \approx  -i (\bdel \omega_c)t\xi \label{number2}
\end{eqnarray}
after omitting a term $(\bdelp a)\exp [-i\omega_c t]$, which may be justified if $a$ varies slowly with $\alpha$ and $\beta$, or because as $t$ increases the term retained on the righthand side of Eq. \ref{number2} dominates. 

We can see how this is consistent with the development of small scales via phase mixing by introducing local wavenumbers for the variation with $\alpha$ and $\beta$,
\begin{eqnarray}
\xi \propto \exp i\left[ \int \kappa_\alpha d\alpha +  \int \kappa_\beta d\beta  \right]. \label{number3}
\end{eqnarray}
Here $\kappa_\alpha$  and $\kappa_\beta$ are the wavenumbers in $\alpha$ and $\beta$ and have units that are the inverse of the units of their respective coordinates. These wavenumbers should be distinguished from the perpendicular components of the usual wave vector {\bf k}, which has units of 1/length. The different wavenumbers may be related through the scale factors ($h$) that relate elemental coordinate increments to physical distances: $d {\bf r} = {\bf e}_\alpha h_\alpha d\alpha  + {\bf e}_\beta h_ \beta d \beta + {\bf e}_\gamma h_ \gamma d \gamma $, where ${\bf e}_\alpha$ is a unit vector in the $\alpha$ direction, etc. In this notation 
$\bdelp = ({\bf e}_\alpha /h_\alpha) \partial /\partial\alpha + ({\bf e}_\beta /h_ \beta) \partial /\partial \beta$, and noting that $\bdelp\xi \approx i {\bf k}_\perp \xi$, eq. \ref{number3} yields
\begin{eqnarray}
\bdelp\xi &=& i {\bf k}_\perp \xi = i\left( \frac{{\bf e}_\alpha}{h_\alpha}\frac{ \partial}{\partial\alpha}
+\frac{{\bf e}_\beta}{h_ \beta}\frac{ \partial}{\partial \beta}\right)\xi \\
&\equiv & i\left( \frac{{\bf e}_\alpha}{h_\alpha}\kappa_\alpha
+\frac{{\bf e}_\beta}{h_ \beta}\kappa_\beta\right)\xi. \label{number4}
\end{eqnarray}
Equating components of the second and fourth expressions in eq. \ref{number4} gives the expected relations between the various wavenumbers,
\begin{eqnarray}
k_\alpha = \kappa_\alpha /h_\alpha, \qquad k_\beta = \kappa_ \beta /h_ \beta. \label{number5}
\end{eqnarray}
Eqs \ref{number2} and \ref{number4} give a direct and elegant expression for the perpendicular wave vector as
\begin{eqnarray}
{\bf k}_\perp \approx -(\bdel \omega_c) t, \label{number6}
\end{eqnarray}
which is a generalisation to three dimensions of the results of
\citet{1995JGR...10019441M}, \citep{1999JGR...10410159W} and \citet{2015ApJ...806..115K} for lower dimensional systems, which developed phase mixing in only one perpendicular coordinate. The above expression allows phase mixing in both perpendicular directions, giving physical phase mixing lengths (or wavelengths) in the $\alpha$ and $\beta$  directions of
\begin{eqnarray}
L_{ph\alpha} = \frac{2\pi}{|k_\alpha |}  \equiv \frac{2\pi h_\alpha}{|\partial \omega_c/\partial\alpha |t}, \qquad
L_{ph\beta} = \frac{2\pi}{|k_ \beta |}  \equiv \frac{2\pi h_\beta}{|\partial \omega_c/\partial \beta |t}. \label{number7}
\end{eqnarray}
If the phase mixing lengths are expressed in the same units as $\alpha$ and $\beta$, rather than physical length as in eq. \ref{number7}, slightly simpler expressions are found, i.e.
\begin{eqnarray}
\ell_{ph\alpha} = \frac{2\pi}{|\kappa_\alpha |}  \equiv \frac{2\pi}{|\partial \omega_c/\partial\alpha |t}, \qquad
\ell_{ph\beta} = \frac{2\pi}{|\kappa_ \beta |}  \equiv \frac{2\pi}{|\partial \omega_c/\partial \beta |t}. \label{number8}
\end{eqnarray}
The development of the phase mixing length can be pictured simply as the tendency for each field line to oscillate with its own natural frequency. Even if all the field lines start to oscillate with the same phase, they soon drift out of phase with one another as time passes. Not only does the phase mixing process generate perpendicular scales, but points of constant phase can be seen to move across field lines. This phase motion has been seen in magnetospheric data of Alfv\'en waves \citep[see the review by][]{2006GMS...169...51W} and the simulations of coronal oscillations by \citet{2015ApJ...806..115K}. These studies note that the direction of motion is related to the spatial variation of $\omega_c$. The results of these papers for the perpendicular phase velocity in physical space generalise to ${\bf V}_{ph} =\omega_c / {\bf k}_\perp$, giving the components
\begin{eqnarray}
V_{ph\alpha} =  \frac{-\omega_c h_\alpha}{(\partial \omega_c/\partial\alpha )t}, \qquad
V_{ph\beta} =  \frac{-\omega_c h_ \beta}{(\partial \omega_c/\partial \beta )t}, \qquad \label{number9}
\end{eqnarray}
If the excitation occurred at a time $t_i$, the subsequent properties are found by replacing $t$ with $t-t_i$ in the above formulae. \\
Even though some of the steps in the above formulation are approximate, the results have been shown to be remarkably robust and valid. For example, Alfv\'en waves are only strictly decoupled when appropriate symmetry is present. Nevertheless, \citet{1995JGR...10019441M} and \citet{2015ApJ...806..115K} show how they can provide an accurate interpretation of simulations which lack this symmetry. Indeed, even when one-dimensional theory is applied to two-dimensional simulations, the expressions work remarkably well \citep{1994JGR....9913455R}.

\subsection{Apparent phase speed}
The generalised expressions for the phase speed that have been derived in both the magnetoshperic and solar literature \citep{1999JGR...10410159W, 2015ApJ...812..121K} can be rewritten for a flux function of $r$ as
\begin{eqnarray}
v_{ph} = \frac{\omega (r)}{-(t-t_i) \frac{\partial \omega (r)}{\partial r}} \label{velocity1},
\end{eqnarray}
with $\omega$ the natural/continuum frequency of the field line in a flux surface at radius $r$. The assumption for writing this formula is that the phase speed (Alfv\'en speed in this case) is a flux function and that the radial coordinate corresponds to the flux coordinate. Adopting the work of \citet{2012ApJ...746...30L}, we get a relation between the angular frequency of an oscillation and the magnetic field,
\begin{eqnarray}
\omega = \sqrt{\frac{g_0}{R_C}} \label{angularfrequency},
\end{eqnarray}
with $g_0$ the gravitational acceleration at the solar surface and $R_C$ the radius of curvature of the fieldline. If we now assume that the radius of curvature $R_C$ is a flux function, then we can use the same formalism as Kaneko et al. (2015) to describe the superslow propagation. To obtain the dependence of the radius of curvature on the height in the prominence, we use the Kippenhahn-Schl\"uter prominence magnetic field model \citep{1957ZA.....43...36K}.
\subsection{Kippenhahn-Schl\"uter model}
\begin{figure}[H]
\begin{center}
\includegraphics[scale=0.8, trim={0 0 0 0}]{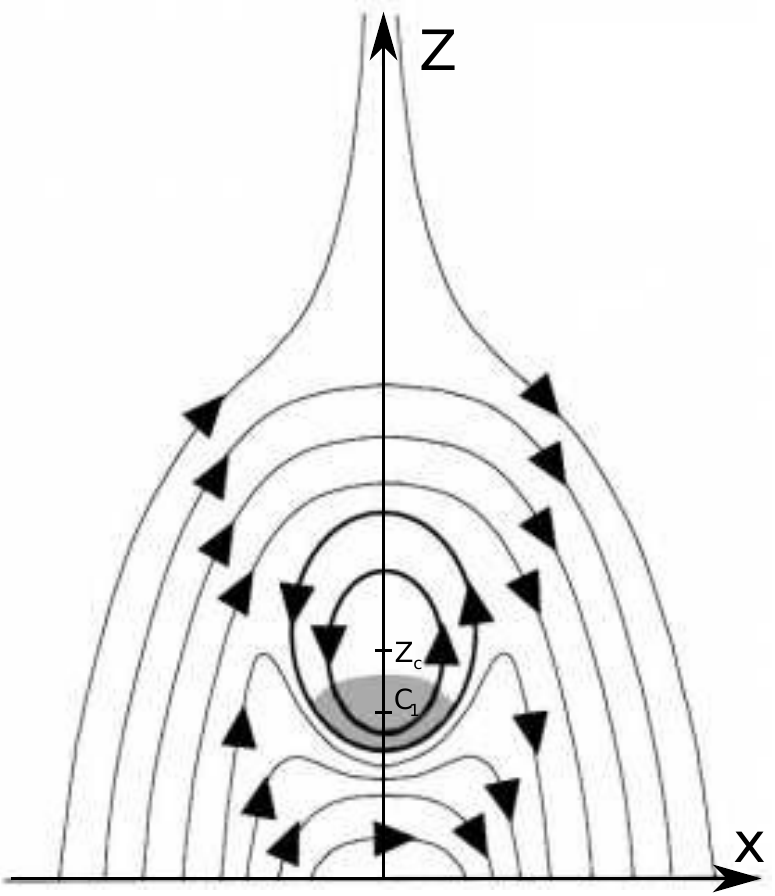} 
\caption{Diagram of the magnetic field configuration of a solar prominence. Figure modified from \citet{2000ApJ...537..503G}. The prominence itself is shown by the shaded area, denoted by $C_1$. This area is modelled by the Kippenhahn-Schl\"uter model.} The value $z_c$ denotes the centre of the magnetic field twist we introduce in section \ref{modified}.
\label{kippenhahnModel}
\end{center}
\end{figure}

The Kippenhahn-Schl\"uter prominence magnetic field model is given by:
\begin{eqnarray}
\begin{cases}
B_x = B_{x0},\\
B_y = B_{y0},\\
B_z(x) = B_{z0} \tanh \Big( \frac{B_{z0}}{2B_{x0}}\frac{x}{\Lambda} \Big), \label{KHmodel}
\end{cases}
\end{eqnarray}
where the x-direction is across the filament, the y-direction is along the filament, and the z-direction is vertical. The quantity $\Lambda$ is the pressure scale height, given by
\begin{eqnarray}
\Lambda=\frac{RT_0}{\tilde{\mu} g}, \nonumber 
\end{eqnarray}
with $R$ the specific gas constant, $T_0$ the temperature (assumed constant), $\tilde{\mu}$ the mean atomic mass, and $g$ the gravitational acceleration. Figure \ref{kippenhahnModel} shows a projection of the xz-plane of the Kippenhahn-Schl\"uter model.
We obtain the equation of the fieldlines in the xz-plane by solving the equation
\begin{eqnarray}
\frac{dx}{B_x} &=& \frac{dz}{B_z},  \\
\int \frac{B_{z0}}{B_{x0}}  \tanh \Big( \frac{B_{z0}}{2B_{x0}}\frac{x}{\Lambda} \Big)dx &=& \int dz,  \\
2 \Lambda \ln \Big\{ \cosh \Big( \frac{B_{z0}}{2B_{x0}}\frac{x}{\Lambda} \Big) \Big\} &=& z+c.  \label{2dfieldline}
\end{eqnarray}
Different values of $c$ give different altitudes in the prominence, which corresponds with the different slices of observations we have, as seen in Figure \ref{allplots}. \\
The radius of curvature of a curve given parametrically by
\begin{eqnarray}
\begin{cases}
x=x(s),  \\
z=z(s), \nonumber
\end{cases} 
\end{eqnarray}
is calculated through
\begin{eqnarray}
R_C = \frac{(x'^2 + z'^2)^{3/2}}{|x'z''-z'x''|}.  \label{radiuscurvature2d} 
\end{eqnarray}
We parametrise the equation of the field line in the Kippenhahn-Schl\"uter model (eq. \ref{2dfieldline}) as
\begin{eqnarray}
\begin{cases}
x=s,  \\
z= 2 \Lambda \ln \Big\{ \cosh \Big( \frac{B_{z0}}{2B_{x0}}\frac{s}{\Lambda} \Big) \Big\} - c. \nonumber 
\end{cases}
\end{eqnarray}
Taking the first and second order derivatives and inserting them into eq. \ref{radiuscurvature2d} gives
\begin{eqnarray}
R_C &=& \frac{ \Big( 1 + \frac{B_{z0}^2}{B_{x0}^2} \tanh^2\Big(\frac{B_{z0}}{2B_{x0}}\frac{s}{\Lambda}\Big) \Big)^{3/2}}{\Big| \frac{B_{z0}^2}{2B_{x0}^2 \Lambda} \Big(1 - \tanh^2\Big(\frac{B_{z0}}{2B_{x0}}\frac{s}{\Lambda}\Big)\Big) \Big|}. \nonumber 
\end{eqnarray}
When considering the centre of the filament ($x=0$), the curvature becomes
\begin{eqnarray}
R_C &=& \frac{1}{\Big|  \frac{B_{z0}^{2}}{2B_{x0}^{2}\Lambda}  \Big|} \nonumber \\
&=&  2\frac{B_{x0}^2}{B_{z0}^2} \frac{RT}{\tilde{\mu} g}.
\end{eqnarray}
This results in a constant value for the radius of curvature, which is contradictory to what we expect. For the concept of apparent superslow waves, we need a radius of curvature that varies with height in the prominence, so that we have a varying angular frequency in the standing waves. We can thus conclude that the assumed model by eq. \ref{KHmodel} is too simplistic for this purpose and we thus introduce a modification.
\subsection{Modified Kippenhahn-Schl\"uter model}
\label{modified}
We modify the model so that the magnetic field in the direction along the prominence depends on the distance to the centre of the prominence, introducing a twist in the magnetic field. We thus take the magnetic field as follows:
\begin{eqnarray}
\begin{cases}
B_x = B_{x0} \label{fieldx}, \\
B_y = \frac{B_{y0}}{S^2}(x^2 + (z-z_c)^2) \label{fieldy}, \\
B_z = B_{z0} \tanh \Big( \frac{B_{z0}}{2B_{x0}}\frac{x}{\Lambda} \Big) \label{fieldz},
\end{cases}
\end{eqnarray}
with $z_c$ the vertical position of the centre of the filament and S a measure for the strength of the magnetic field twist. To get the radius of curvature in this three-dimensional scenario, we use a slightly different approach, as we only need the derivative of the field line parametrisation. A field line with parametric equation $\textbf{r}(s)$ must have its tangent vector $d\textbf{r}/ds$ parallel to $\textbf{B} (\textbf{r} (s))$. This means that
\begin{eqnarray}
\frac{d\textbf{r}}{ds} = \lambda(s) \textbf{B}(\textbf{r} (s)). \label{differntialVec}
\end{eqnarray}
The differential equations for the fieldlines are then given by
\begin{eqnarray}
\frac{dx}{B_x} = \frac{dy}{B_y} = \frac{dz}{B_z}. \nonumber 
\end{eqnarray}
Solving $\frac{dx}{B_x} = \frac{dz}{B_z}$ gives us the same result as eq. \ref{2dfieldline}. We then introduce a parametrisation as before as follows:
\begin{eqnarray}
\begin{cases}
x=s, \\
z= 2 \Lambda \ln \Big\{ \cosh \Big( \frac{B_{z0}}{2B_{x0}}\frac{s}{\Lambda} \Big) \Big\} + C_1.
\end{cases}
\end{eqnarray}
Differentiation yields
\begin{eqnarray}
x'&=&1, \label{firstdiffx} \\
z' &=&  \frac{B_{z0}}{B_{x0}} \tanh \Big(\frac{B_{z0}}{2B_{x0}}\frac{s}{\Lambda} \Big). \label{firstdiffz}
\end{eqnarray}
Using eq. \ref{firstdiffx} with eqs \ref{fieldx} and \ref{differntialVec} yields a value for $\lambda(s)$,
\begin{eqnarray}
\lambda (s) = \frac{1}{B_{x0}}.
\end{eqnarray}
Using eq. \ref{firstdiffz} with eqs \ref{fieldz} and \ref{differntialVec} confirms this value for $\lambda(s)$. Combining eqs \ref{fieldy} and \ref{differntialVec} with $\lambda(s)$ then gives us
\begin{eqnarray}
y'&=& \frac{B_{y0}}{S^2B_{x0}} (x^2 + (z-z_c)^2) \nonumber  \\
&=& \frac{B_{y0}}{S^2B_{x0}} (s^2 + (2\Lambda \ln \Big\{ \cosh \Big( \frac{B_{z0}}{2B_{x0}}\frac{s}{\Lambda} \Big) \Big\} \nonumber \\
&+& C_1-z_c)^2).
\end{eqnarray}
After calculating the second order derivatives, the radius of curvature for a three-dimensional fieldline in the centre of the prominence can then be calculated as follows:
\begin{eqnarray}
R_c &=& \frac{(x'^2+y'^2 + z'^2)^{3/2}}{\sqrt{(z''y'-y''z')^2 + (x''y'-y''x')^2 + (x''z'-z''x')^2}} \nonumber  \\
 &=& \frac{2 \Lambda B_{x0}^2}{B_{z0}^2} \Big(1+\frac{B_{y0}^2}{B_{x0}^2 S^4}(C_1-z_c)^4 \Big) \label{radcurve3D}
\end{eqnarray}
Rewriting this equation using
\begin{eqnarray}
v &=& B_{y0}/B_{x0}, \label{v_substi}  \\
w &=& B_{z0}/B_{x0}, \label{w_substi}  \\
A &=& \frac{C_1-z_c}{S},  \label{A_substi} 
\end{eqnarray}
gives us
\begin{eqnarray}
R_c = \frac{2 \Lambda}{w^2} (1+v^2 A^4).
\end{eqnarray}
\subsection{Apparent phase velocity}
Applying this result to the formula for the angular frequency of \citet{2012ApJ...746...30L} (eq. \ref{angularfrequency}) gives us
\begin{eqnarray}
\omega &=& \sqrt{\frac{g}{R_c}} \nonumber \\
&=& \sqrt{\frac{g w^2}{2\Lambda}\frac{1}{1+v^2 A^4}},
\end{eqnarray}
which is a function of height. 
Before we can use this, we need to calculate the $\frac{\partial \omega (r)}{\partial r}$ term in eq. \ref{velocity1}. We can rewrite this as
\begin{eqnarray}
\frac{\partial \omega (r)}{\partial r} &=& \frac{\partial \omega}{\partial A} \frac{\partial A}{\partial r} \vec{r},
\end{eqnarray}
where $A$ is as we defined in eq. \ref{A_substi}. The $r$ coordinate here is the same as our $C_1$ coordinate, making the $\frac{\partial A}{\partial r}$ factor equal $\frac{1}{S}$. This then gives
\begin{eqnarray}
v_{ph} = -\frac{S}{t-t_i} \frac{\omega}{\frac{\partial \omega}{\partial A}}
\end{eqnarray}
Calculating the derivative of the angular frequency yields
\begin{eqnarray}
\frac{\partial \omega}{\partial A} &=& \frac{1}{2\sqrt{\frac{g w^2}{2\Lambda}\frac{1}{1+v^2 A^4}}} \Big( -\frac{g w^2}{2\Lambda} \frac{1}{(1+v^2 A^4)^2}4v^2A^3  \Big) \nonumber \\
&=& - \frac{2 \omega v^2 A^3}{1+v^2A^4}.
\end{eqnarray}
Inserting this into the phase velocity equation gives us
\begin{eqnarray}
v_{ph} 
 &=& \frac{S}{t-t_i}  \frac{1+v^2A^4}{2 v^2 A^3} \label{finalvelocity}
\end{eqnarray}
A first thing to notice is that the $w$ quantity that describes the ratio of $B_{z0}$ to $B_{x0}$ vanishes completely, meaning that the apparent phase velocity is independent of the scale of the magnetic field in the vertical direction. It only depends on time, guide field, and height. 
Assuming that $A$ in eq. \ref{finalvelocity} is large compared to the $v$ component, we can rewrite the equation so that
\begin{eqnarray}
v_{ph} &=& \frac{S}{t-t_i} \frac{A^4(\frac{1}{A^4} + v^2)}{A^3 2v^2} \nonumber \\
&\approx &  \frac{S}{t-t_i} \frac{A^4}{A^3} \frac{v^2}{2v^2} \nonumber \\
&=&  \frac{S}{t-t_i} \frac{1}{2} A \nonumber \\
&=& \frac{1}{2} \frac{1}{t-t_i} (C_1-z_c)
\end{eqnarray}
In this limit for large $A$
(corresponding to a large flux rope, with a slowly varying twist), the parameter for the magnetic twist magnitude $v$ is no longer present in the equation. Surprisingly, the phase speed of the apparent superslow propagation only depends on the distance to the centre of the flux rope.
\subsection{Fitting the observations}
\label{fitobserv}
We thus fit our three observed phase velocities (14, 8, and 4 km/s) to their observed times ($t$ equals 1390, 5580, and 9501 s) in order to get values for $t_i$ and $C_1-z_C$. Doing so yields a value of $-2170$ s for $t_i$ and $103$ Mm for $C_1-z_c$. A plot of this fit can be found in figure \ref{velocitieshigh}.
An excitation time of $2170$ s roughly equals one apparent oscillation period. This means that the time of excitation of our wave happened about one oscillation period before we observed the end of the eruption, thus around 02:45 UT. From a physical point of view, $C_1-z_c$ basically gives the distance from the centre of the magnetic field twist to the filament. According to various works and observations,  the prominence itself is located low in the dip of the magnetic field lines, also called a cavity (as can be seen in Figure \ref{kippenhahnModel}). This more or less circular structure can reach up to twice the height of the prominence itself and extends well above its top \citep{2006JGRA..11112103G, 2010ApJ...724.1133G, 2010SSRv..151..333M, 2012ApJ...748L..26X}. Looking at the fact that the filament height can reach up to $10^5$ km, the obtained value for the prominence cavity size is compatible with the earlier observations. We tried using other observations from AIA and SDO at later times to confirm our results about the height. This proved to be fruitless however, as no observations are clear enough for a proper sign of the flux rope or cavity.
since our value for $C_1-z_c$ is positive, we can deduce that the centre of the magnetic field twist is located to the left of our slices in Figure \ref{prominencefigures}. This can be confirmed by looking at the oscillation frequency of the different slices. Figure \ref{allplots} shows us that the oscillation frequency decreases when moving through the slices from left (height 0) to right (height 7637 km). Eq. \ref{angularfrequency} tells us that decreasing frequency corresponds with increasing radius of curvature, which in our modified Kippenhahn-Schl\"uter model conforms to moving away from the centre of magnetic field twist.

\subsection{Alfv\'en and slow waves}
The longitudinal oscillations described by \citet{2012ApJ...746...30L} are not the only modes that can create the effect of apparent waves. When using Alfv\'en waves for this purpose, the angular frequency is given by
\begin{eqnarray}
\omega &=& k v_a \nonumber \\
&=& k \sqrt{\frac{B^2}{\mu \rho}}.
\end{eqnarray}
We take the wavenumber $k$ and the density $\rho $ to be constant. Using the modified Kippenhahn-Schl\"uter model (eq. \ref{fieldx}), we get at the centre of the filament
\begin{eqnarray}
B &=& \sqrt{B_{x0}^{2} + \frac{B_{y0}^{2}}{S^4}(z-z_c)^4}.
\end{eqnarray}
The r coordinate in eq. \ref{velocity1} is the same as the z coordinate in this situation. We thus calculate the derivative of the angular frequency as follows:
\begin{eqnarray}
\frac{\partial \omega}{\partial z} 
&=& \frac{k}{\sqrt{\mu \rho}} \frac{2 \frac{B_{y0}^{2}}{S^4}(z-z_c)^3}{\sqrt{B_{x0}^{2} + \frac{B_{y0}^{2}}{S^4}(z-z_c)^4}}.
\end{eqnarray}
The apparent phase velocity then becomes
\begin{eqnarray}
v_{ph} &=& -\frac{1}{t-t_i} \frac{\omega}{\frac{\partial \omega}{\partial z}} \\
&=& \frac{\frac{k}{\sqrt{\mu \rho}} \sqrt{B_{x0}^2 + \frac{B_{y0}^{2}}{S^4}(z-z_c)^4}}{\frac{k}{\sqrt{\mu \rho}} \frac{2 \frac{B_{y0}^{2}}{S^4}(z-z_c)^3}{\sqrt{B_{x0}^{2} + \frac{B_{y0}^{2}}{S^4}(z-z_c)^4}}} \\
&=& S \frac{1 + \frac{B_{y0}^{2}}{B_{x0}^{2}}\frac{(z-z_c)^4}{S^4}}{2 \frac{B_{y0}^{2}}{B_{x0}^{2}} \frac{(z-z_c)^3}{S^3}}.
\end{eqnarray}
Rewriting this using eq. \ref{v_substi} to \ref{A_substi} gives the following:
\begin{eqnarray}
v_{ph} &=& \frac{S}{t-t_i} \frac{1+v^2 A^4}{2 v^2 A^3}.
\end{eqnarray}
It turns out that with the exception of the minus sign, the apparent phase velocity when using Alfv\'en waves is the same as when using the gravity waves. This can be attributed to the fact that in our modified Kippenhahn-Schl\"uter model, the radius of curvature is proportional to $B^{-2}$. When using these quantities for the angular frequencies, and thus apparent phase velocity, these similarities result in near identical outcomes. In any case, Alfv\'en waves have displacements perpendicular to the field and this does not seem to be compatible with the observations. \\
We could also consider slow waves for the purpose of apparent superslow waves. However, for this to be valid, we need a smooth transition in temperature. This is not the case, as there is a fast variation in temperature between the filament core and surrounding corona \citep{2012ApJ...748L..26X, 2009ApJ...699.1553S}. Therefore, slow waves have not been considered for this purpose.

\section{Conclusions}
Oscillatory motion has been detected in the 2015 March 15 prominence observed with SDO/AIA (193 \r{A}). Data reduction was performed to properly locate the edge of the prominence. This allowed us to measure the amplitude of the oscillation across a set of slices. From this we derived the velocity of the propagation for three separate wave-like motions; these speeds are 14, 8, and 4 km/s. Both the low values and the presence of the decrease over time shows that this motion cannot be interpreted as MHD waves. Instead, we suggest that this is evidence for superslow propagation,  which is an illusionary effect created by phase mixing of standing and/or slow waves trapped in closed magnetic structures in the prominence. The case we studied is possibly not an exception, but it could very well be that apparent superslow waves occur rather often in solar prominences.\\
We have generalised the concept of superslow waves to three dimensions and extended it to gravity waves in prominences, as proposed by \citet{2012ApJ...750L...1L}, where we have assumed that the radius of curvature of the prominence field lines is a flux function.\\
When using the measurements for seismology we can conclude that the Kippenhahn-Schl\"uter model is too simple to get any results, as the radius of curvature is constant everywhere. Using a modified model including a spatially varying guide-field we derive the dependence of the radius of curvature with height. We can conclude that the scale of the magnetic field in the vertical direction plays no role in the concept of apparent superslow waves. Fitting our formula for the apparent speed in the superslow propagation to the data, we learn that the moment of excitation happens roughly one oscillation period before the end of the eruption and we obtain a value of $103$ Mm for the distance between the filament and flux rope
axis. Thus, for the first time, we have performed seismology of superslow propagating waves to characterise the magnetic structure of a prominence.

\bibliographystyle{aa}
\bibliography{superslowmanuscript}

\end{document}